\documentclass[sigconf]{acmart}

\usepackage{booktabs} 

\acmYear{2017} 
\setcopyright{rightsretained} 
\acmConference{WebSci '17}{June 25-28, 2017}{Troy, NY, USA}
\acmDOI{10.1145/3091478.3098865}
\acmISBN{978-1-4503-4896-6/17/06}

\begin{document}

\title{Predicting Research that will be Cited in Policy Documents}

\author{Bharat Kale}
\affiliation{%
	\institution{Northern Illinois University}
}
\email{bkale@niu.edu}

\author{Harish Varma Siravuri}
\affiliation{%
	\institution{Northern Illinois University}
}
\email{hsiravuri@niu.edu}
       
\author{Hamed Alhoori}
\affiliation{%
	\institution{Northern Illinois University}
}
\email{alhoori@niu.edu}

\author{Michael E. Papka}
\affiliation{%
	\institution{Argonne National Laboratory}
    \institution{Northern Illinois University}
}
\email{papka@niu.edu}

\renewcommand{\shortauthors}{B. Kale et al.}

\begin{abstract}

Scientific publications and other genres of research output are increasingly being cited in policy documents. Citations in documents of this nature could be considered a critical indicator of the significance and societal impact of the research output. In this study, we built classification models that predict whether a particular research work is likely to be cited in a public policy document based on the attention it received online, primarily on social media platforms. We evaluated the classifiers based on their accuracy, precision, and recall values. We found that Random Forest and Multinomial Naive Bayes classifiers performed better overall.

\end{abstract}

%
%

\keywords{Public Policy, Policy documents, Altmetrics, Social Media}


\maketitle

\section{Introduction and Related Work}
Policy documents influence large sections of society \cite{Bornmann2016}. Because of the unique importance of policy documents across diverse organizations, citations included in this type of material support both the credibility of the author cited and the credibility of the policy document itself \cite{freeman2011documents}. Likewise, in this context, it may be appropriate to assign a policy document citation more weight than a regular citation included in a literature review in a scholarly paper, for example. 

Haunschild and Bornmann \cite{Haunschild2017} studied the percentage of papers in Web of Science that are mentioned in policy-related documents and found that less than 0.5\% of the papers on a range of subjects had been mentioned at least once in policy-related documents. Lauren \cite{cadwallader2016papers} analyzed patterns in the types of altmetric attention received by papers that make it into policy documents and found that papers are often being referenced quickly, i.e., within 2 years of publication, such that they are having a real-world impact sooner than expected

Winterfeldt \cite{Winterfeldt20082013} presented a framework to bridge the gap between science and decision making in the policy sphere. Orduna-Malea, Thelwall, and Kousha [7] explored the relationship between citations in patents and technological impact and found that the number of patents citing a resource indicates the technological capacity or relevance of that resource. According to Black \cite{Black275}, although evidence-based policy-making is being encouraged in all areas of public service, research is currently under-used in policy-making and there is a need for a better mutual  understanding between research and policy communities.

Citation analysis is self-limiting because it does not account for many other signals through which research receives attention. An increasing amount of scholarly content is being shared and discussed daily on social media platforms \cite{Euan}. Whereas citations measure research impact within scholarly boundaries, non-traditional web-based metrics or altmetrics \cite{holmberg2015altmetrics}\cite{alhoori2014altmetrics} make it possible to measure different influences, including readers who read an article or share, and/or discuss it with others, but do not formally cite it in traditionally published articles.

Thelwall et al. \cite{thelwall2016alternative} studied the potential value of altmetrics for evaluating funding criteria and found that some metrics can be helpful in this sphere. Sarewitz and Pielke \cite{sarewitz2007neglected} proposed a method to strengthen the connection between science policy decisions, scientific research, and social outcomes using the example of climate change research. Pawson \cite{pawson2002evidence} discussed various ways to incorporate research results into the policy-making process. To date, most studies focus on understanding and using altmetrics in reference to only a few measures. The present study is the first to explore modeling altmetrics in order to predict citations in policy documents.

\section{Data Collection}
The dataset in this study is a database dump that we obtained from altmetric.com, which consists of 5.2 million articles. Our initial analysis showed that of these articles, 89,350 had been cited in at least one policy document whereas 5,097,207 had not been included in a document of this kind. To create a balanced dataset for further analysis, along with the 89,350 articles that had been cited in a policy document, we randomly chose another 89,350 articles that had not been cited in a policy document. The result was a balanced dataset with approximately 180,000 records, half of which had been cited in policy documents.

\section{Feature Selection}
The dataset has a very rich set of features for each article. However, in our analysis, we considered only  features related to online attention. The dataset consists of mention counts on various online sources including reference managers, mainstream news outlets, blogs, peer-review platforms (e.g., PubPeer and Publons), social media, public policy documents, and Wikipedia. 

We used mention counts on Twitter, Facebook, Reddit, Mendeley, Google+, Wikipedia, Weibo, mainstream news outlets, blogs, videos, and peer review sites as features to build the classifiers. Yet, we left a few sources out of our account, including Connotea, which was discontinued in 2013, and Pinterest and Stackoverflow, which together contributed to less than 1\% of the articles in the sample. We replaced the policy citation count with a binary class label denoting whether a given article had been cited in a policy document.

\section{Methods and Results}
\subsection{Classification}
To predict the likelihood of a research article being cited in a policy document, we implemented three classifiers: the Multinomial Naive Bayes classifier, the Random Forest classifier with the number of trees set at 100, and a C-Support Vector Machine with the Radial Basis Function (RBF) kernel. We then divided the entire dataset into training and test sets comprising 80\% and 20\% of the entire dataset, respectively. We trained the models using a 10-fold cross-validation technique and evaluated them based on accuracy, precision, recall, and F1-measure metrics, as shown in Table 1. 

\begin{table}[!h]
\centering
\caption{Accuracy, Precision, Recall, and F1-Measure for different models}
\begin{tabular}{|l|c|c|c|} 
\hline
&Multinomial Naive Bayes&Random Forest&SVM\\ \hline
Accuracy & $0.842$ & $0.870$ & $0.868$\\ \hline
Precision & $0.802$ & $0.826$ & $0.820$\\ \hline
Recall & $0.905$ & $0.870$ & $0.868$\\ \hline
F1-Measure & $0.850$ & $0.844$ & $0.824$\\ \hline
\end{tabular}
\end{table}

\subsection{Feature Ranking}
With the classification models built, we calculated the weight for each feature to determine the significance of each in making the final prediction. Given that feature weights in the case of a Support Vector Machine can be determined only for linear kernels, we ranked the features based on their relevance for only the Random Forest and Multinomial Naive Bayes classifiers. We ranked the features in regard to their importance to the Random Forest classifier from most to least important, as shown in Table 2.

\begin{table}[!h]
\centering
\caption{Feature ranking for different models}
\begin{tabular}{|l|c|c|} \hline
Platform&Random Forest&Multinomial Naive Bayes\\ \hline
peer-review & $0.273595$ & $4.4267$\\ \hline
Google+ & $0.197488$ & $3.4210$\\ \hline
Reddit & $0.151016$ & $4.4087$\\ \hline
video & $0.098035$ & $4.9458$ \\ \hline
Twitter & $0.068745$ & $2.2421$ \\ \hline
Weibo & $0.088242$ & $3.7988$\\ \hline
Mendeley & $0.030116$ & $0.3210$ \\ \hline
Wikipedia & $0.026027$ & $4.9668$ \\ \hline
blogs & $0.018631$ & $4.4571$ \\ \hline
Facebook & $0.016189$ & $3.2314$ \\ \hline
news & $0.008926$ & $3.7307$ \\
\hline\end{tabular}
\end{table}

\section{Conclusions And Future Work}
In this study, we used a specific set of features that track online attention received by scholarly articles to build classifiers to predict the likelihood of an article being cited in public policy. The Random Forest classifier showed better results in making predictions.  We found mention counts on peer-review platforms to be the most influential feature, whereas news rated as the least influential feature. The promising results obtained in this work show that a relationship exists between the online attention that a scholarly work receives and the policy citations it generates, which we were able to exploit. We intend to extend our work in this area by building regression models to predict the number of policy citations a given work is likely to receive. We also plan to build more classifiers with different feature sets and to compare our results.

\begin{acks}
MEP was supported in part by the Office of Advanced Scientific Computing Research, Office of Science, U.S. Department of Energy, under Contract DE-AC02-06CH11357.
\end{acks}

\bibliographystyle{ACM-Reference-Format}
\bibliography{sigproc} 

\end{document}